\documentclass[aps,prb,reprint,showkeys,showpacs]{revtex4-2}
\pdfoutput=1
\usepackage{microtype}
\usepackage{amsmath}
\usepackage{amssymb}
\usepackage{amsfonts}
\usepackage{graphicx}
\usepackage{bm}
\usepackage{siunitx}
\usepackage{color}
\usepackage{hyperref}
\hypersetup{%
    pdfborder = {0 0 0}
}

\allowdisplaybreaks
\usepackage{cleveref}
\usepackage[version=4]{mhchem}

\graphicspath{{./figures/}}

\renewcommand\vec{\bm} % denote vectors with bold
\newcommand{\uGrad}{\vec{\nabla}}

\newcommand{\ud}{\,{\mathrm{d}}}

\newcommand{\uiiint}{\int\!\!\!\int\!\!\!\int}

\newcommand{\uC}{C} % exchange stiffness C=2A
\newcommand{\uD}{D} % Dzyaloshinskii-Moriya constant
\newcommand{\ue}{e} % trial function e(x)
\newcommand{\uE}{E} % energy per unit volume
\newcommand{\uEE}{\mathcal{E}} % total hopfion energy
\newcommand{\uHopf}{{\cal{H}}} % Hopf index
\newcommand{\uh}{h} % dimensionless magnetic field
\newcommand{\uH}{H} % ext. magnetic field
\newcommand{\uvH}{\vec{H}}     % -''-''- vector
\newcommand{\uvHD}{\vec{H}_\mathrm{D}} % demag. field
\newcommand{\uK}{K}         % anisotropy constant
\newcommand{\uvM}{\vec{M}}
\newcommand{\uMi}{M_i}      % i=X,Y,Z
\newcommand{\uvm}{\vec{m}}
\newcommand{\umz}{m_{\mathrm{Z}}}

\newcommand{\uO}{O}         % Cartesian system origin
\newcommand{\uP}{P_i}
\newcommand{\uPex}{P_\mathrm{EX}}
\newcommand{\uPa}{P_\mathrm{A}}
\newcommand{\uPz}{P_\mathrm{Z}}
\newcommand{\uPdm}{P_\mathrm{DM}}
\newcommand{\uPms}{P_\mathrm{MS}}
\newcommand{\up}{p_i}
\newcommand{\upex}{p_\mathrm{EX}}
\newcommand{\upa}{p_\mathrm{A}}
\newcommand{\upz}{p_\mathrm{Z}}
\newcommand{\updm}{p_\mathrm{DM}}
\newcommand{\upms}{p_\mathrm{MS}}
\newcommand{\uetaeq}{\eta_\mathrm{EQ}}
\newcommand{\uEEeq}{\uEE_\mathrm{EQ}}
\newcommand{\uMs}{M_{\mathrm{S}}}
\newcommand{\uMssq}{M_{\mathrm{S}}^2}
\newcommand{\umuZ}{\mu_0}
\newcommand{\uq}{q} % dim.less anistoropy in units ~D
\newcommand{\uvs}{\vec{s}}  % global anisotropy axis
\newcommand{\uX}{X}         % Cartesian X coordinate
\newcommand{\uY}{Y}         % Cartesian Y coordinate
\newcommand{\uZ}{Z}         % Cartesian Z coordinate
\newcommand{\ur}{r}

\newcommand{\uvr}{\vec{r}}          % radius-vector
\newcommand{\uvrp}{\vec{r}^\prime}  % -''- primed
\newcommand{\ut}{\theta}
\newcommand{\uf}{\varphi}

\begin{document}
\title{Spatial decay of isolated magnetic hopfions in chiral helimagnets}

\author{Konstantin L. Metlov}\email{metlov@donfti.ru}
\affiliation{Galkin Donetsk Institute for Physics and Engineering, R.~Luxembourg str.~72, Donetsk 283048, Russian Federation}

\author{Maksim M. Gordei}
\affiliation{Galkin Donetsk Institute for Physics and Engineering, R.~Luxembourg str.~72, Donetsk 283048, Russian Federation}

\date{\today}
\begin{abstract}
Magnetic hopfions in chiral magnets are topological solitons, localized in three dimensions. But is their localization strong? To address this question we derive an asymptotic expansion for the isolated hopfion's spatial profile. It becomes starting point for a simple analytical model, which is asymptotically correct both near the hopfion center and far away from it. Region of equilibrium hopfions on the phase diagram of a helimagnet is computed and material requirements for supporting movable isolated magnetic hopfions on uniform background are discussed.
\end{abstract}


\maketitle

\section{Introduction}
In magnetically ordered media (ferromagnets, ferrimagnets, antiferromagnets, etc) the field of vectorial order parameter (such as magnetization, or antiferromagnetism vector) can host a number of localized topological objects (solitons) of various dimensions --- (one-dimensional) domain walls~\cite{AharoniBook,Hubert_book_walls}, (two-dimensional) skyrmions~\cite{RBP2006} and magnetic vortices~\cite{UP93} or (three-dimensional) hopfions~\cite{DI79} and Bloch point~\cite{Feldtkeller1965, doring68} assemblies with compensated topological charges. But how strongly are these objects localized? What is their spatial extent? For example, it is a very well known from the theory of domain walls (see e.g. Chapter 8 of Ref.~\onlinecite{AharoniBook}) that N{\'e}el domain wall's profile is best described by Ditze and Thomas variational model~\cite{DT61} with polynomial approach to a constant direction away from the wall, while Bloch wall is fitted better by Landau and Lifshitz model~\cite{LL35}, whose non-uniformity decays exponentially. It is often said in this regard that N{\'e}el walls have ``long tails''. Skyrmions have exponentially decaying ``short'' tails, proportional to the modified cylindrical Bessel's function~\cite{IS86,Tiwari2019}, while the tails of magnetic vortices are long and oscillating~\cite{GN04_JAP_PERP}.

The central question of the present work is --- what kind of ``tails'' magnetic hopfions in a helimagnet have and how to control them? This question also has practical implications. Complex magnetization states in three dimensions (3D) can be described (up to a homotopy at least) via products of quaternionic functions~\cite{M2025quaternionic}. However, the algebra of quaternions (just like the algebra of 3D rotation matrices) is non-commutative, or, in other words --- non-Abelian. It implies symmetry breaking in the magnetization vector field during pairwise interaction of hopfions~\cite{M2025quaternionic}. This non-Abelianness (also shared by Majorana fermions) is a unique feature of the emerging 3D spintronics~\cite{Gubbiotti2024roadmap}, which has no direct two-dimensional (2D) counterpart. This is because 2D magnetization configurations can be described as products of functions of complex variable~\cite{doring68,BP75,M10} and multiplication in the complex algebra is {\em commutative}. Thus, interacting hopfions can be instrumental for implementing topological quantum computation, based on the algebra of braids, as was originally proposed by Alexei Kitaev~\cite{Kitaev2003}.

The hopfion tails are key to realizing such non-Abelian interactions. If the tails are short and non-overlapping, the hopfion braiding has no physical effect on the magnetization distribution. The longer are the tails, the larger group of hopfions can be brought into a strongly interacting (entangled) state with broken symmetry, reflecting history of their interchanges.

Here we study variational model of an isolated hopfion (with the Hopf index $\uHopf=1$) on the uniform background, which is a generalization~\cite{M2025quaternionic} of the strongly localized hopfion model in a lattice~\cite{M2025} and is homotopically equivalent (as all the hopfions are) to the original Whitehead's ansatz~\cite{whitehead1947}. Magnetization rotation within the hopfion is parametrized via the so-called profile function $\ue(t)$, where $t$ is the normalized distance from the hopfion center. We find and discuss an asymptotic solution for hopfion tails $\ue(t\gg1)$. It guides us to a simple model for the profile function, which is asymptotically correct both at small and at large $t$ and, at the same time, simple enough to compute the equilibrium hopfion energy and size analytically in the closed form. We identify the region of the magnetic phase diagram of an anisotropic helimagnet, where isolated magnetic hopfions on uniform background can be realized.

\section{Hopfion energy}

Let us start with classical micromagnetic energy density of an infinite 3D helimagnet, containing the magnetization distribution $\uvM(\uvr)$:
\begin{equation}
\label{eq:energy} 
\begin{aligned}
        \uE = & \frac{\uC}{2 \uMssq}\sum_{i=\uX,\uY,\uZ} \left|\uGrad\uMi\right|^2 +\frac{\uD}{\uMssq}\, \uvM\cdot[\uGrad\times\uvM] -\\
        &   -\umuZ \left(\uvM \cdot \uvH\right) - \frac{\umuZ}{2}\left(\uvM \cdot \uvHD\right)- \frac{\uK}{\uMssq}\left(\uvM \cdot \uvs\right)^2,
\end{aligned}
\end{equation}
where $C=2A$ is the exchange stiffness; $\uMs=|\uvM(\uvr)|$ is the saturation magnetization; $\uvr=\{\uX,\uY,\uZ\}$; $\uX$,$\uY$ and $\uZ$ are the coordinates of the chosen Cartesian coordinate system; $D$ is the Dzyaloshinskii-Moriya interaction strength; $\umuZ$ is the permeability of vacuum; $\uvH$ is the external magnetic field; $\uvHD$ is the demagnetizing field, created by the whole $\uvM(\uvr)$ as per Maxwell's equations; $\uK$ and $\uvs$ are the uniaxial anisotropy constant and director. We assume that directions of $\uvH=\{0,0,\uH\}$ and $\uvs=\{0,0,1\}$ coincide and are along the $\uO\uZ$ axis of the coordinate system. The hopfion is located at $\uvr=0$, surrounded by (an almost) uniform background such that far away at infinity  the magnetization is also directed along the $\uO\uZ$ axis: $\uvm(\uvr)_{|\uvr|\rightarrow\infty}\rightarrow\{0,0,1\}$.

\begin{figure}[t]
\begin{center}
\includegraphics[width=\columnwidth]{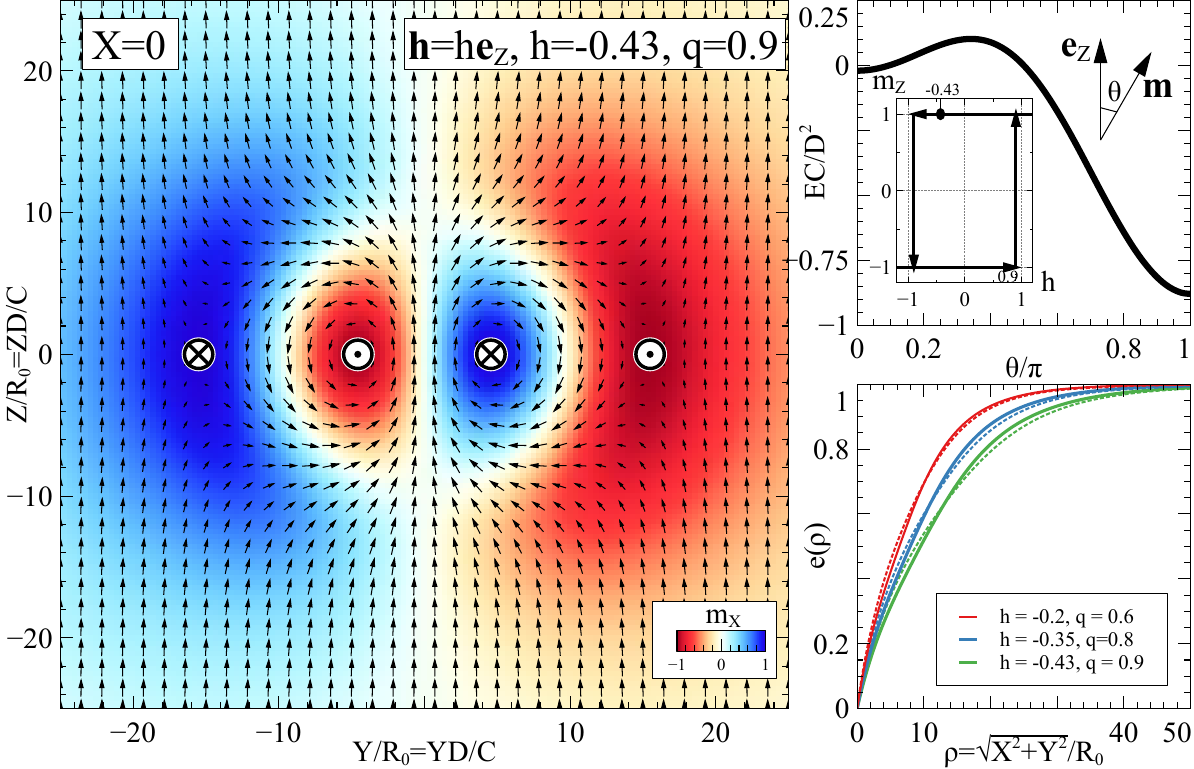}
\end{center}
\caption{\label{fig:hopfion}(left) Cross section of the equilibrium hopfion~\eqref{eq:ansatz} by the $X=0$ plane through its central $\uO\uZ$ axis; (top right) the energy of a uniform magnetization background at an angle $\theta$ to the anisotropy axis, inset shows the saturation-to-saturation hysteresis loop, where dot marks the field, experienced by the hopfion on the left; (bottom right) equilibrium hopfion profile functions from the Eq.~\ref{eq:trialE} (solid) and numerically computed (dotted). The dimensionless magnetic field $h$ and the anisotropy quality factor $q$ are defined in the text.}
\end{figure}

The second starting point is the quaternionic ansatz~\cite{M2025quaternionic} for the magnetization distribution $\uvM=\uMs \uvm$ within a hopfion, which in the spherical coordinate system $\{\rho,\varphi,\theta\}$ for $\uvm=\{m_\rho, m_\varphi,m_\theta\}$ reads
\begin{align}
 \label{eq:ansatz}
 \uvm=&\left\{\cos\theta,\frac{4g(2e-1)\sin\theta}{(1-2g)^2},\frac{(4g(1+g)-1)\sin\theta}{(1-2g)^2}\right\},
\end{align}
where $g=e(1-e)$, $e=e(\rho/R)$ and $R$ is related to the (yet unknown) hopfion radius. There can be many different conventions on how to define the hopfion radius, but all of them can be expressed as a certain dimensionless prefactor times $R$. The power of the quaternionic ansatz lies in the possibility to represent many hopfions on top of systems of Bloch points and other magnetization backgrounds. For a single hopfion it coincides (sans the homotopy, defined by the $e(t)$ profile function) with the original Whitehead's ansatz~\cite{whitehead1947}. Yet, the expression~\eqref{eq:ansatz} is the exact rendering of the quaternionic ansatz~\cite{M2025quaternionic} for a single $\uHopf=1$ hopfion, while representing the scaling factor~\cite{M2025quaternionic} $f(t)=[1/t]e(t)/[1-e(t)]$. Such a hopfion is axially symmetric (around the $\uO\uZ$ axis) and its cross-section is shown in Fig.~\ref{fig:hopfion}.

Substituting~\eqref{eq:ansatz} into~\eqref{eq:energy} we can express normalized hopfion energy $\uEE=(D/C^2)\int_V E \ud^3\uvr = \int_0^\infty F \ud t$ with
\begin{align}
\label{eq:Penergy}
 F  = & \eta \uPex + \eta^2 \uPdm + \eta^3 (h \uPz + \frac{q}{2} \uPa + \mu^2 \uPms),
\end{align}
where we have introduced dimensionless hopfion radius $\eta = R/R_0$, with $R_0=C/D$. The other normalized parameters are the magnetic field $h=\umuZ \uMs C H/D^2$, the anisotropy quality factor $q=2CK/D^2$ and the magnetostatic interaction strength $\mu^2 = \umuZ\uMssq C/D^2$ (also known as the susceptibility of the conical helix~\cite{GWG2017}). This normalization assumes the presence of Dzyaloshinskii-Moriya interaction ($D\ne0$). The quantities $\uP$ are given in the Appendix~\ref{app:Pfuncs}. They depend on $t$, the profile function $e$ and its first derivative. Due to non-locality of the dipolar interaction, $\uPms$ contains another integral over $t$. The functions $\uPz$ and $\uPa$ have additional constant terms, chosen in such a way that the energy of the background state turns out to be exactly $0$, so that~\eqref{eq:Penergy} represents the energy, gained by placing a single hopfion at the origin of the coordinate system in an otherwise uniformly magnetized bulk magnet.

There are two differences, compared to the previously studied model of the infinite densely-packed hopfion lattice~\cite{M2025}. First is minor --- there is no lattice cell and the integration in~\eqref{eq:Penergy} goes across the whole space (with angular integrals inside the $\uP$). The second is major --- the energy of the hopfion lattice is infinite and minimizing it implies minimization of the energy density {\em per unit cell}, which means that the energy of the hopfion must be additionally divided by the lattice cell size, roughly $\propto R^3$. The isolated hopfion, considered here, has finite energy, which is minimized directly, without any renormalization. This difference boils down to an additional ``lattice pressure'', which is experienced by hopfions in the lattice and is absent for an isolated hopfion. That's why (similarly to the cases of magnetic bubble lattices~\cite{DD71} and isolated magnetic bubbles~\cite{Thiele:1969:TCM}) the energetic aspects (equilibrium profile, size, energy) of isolated hopfions are different from those in the lattice. One can directly see the effect of the lattice pressure by comparing the Fig.~\ref{fig:hopfion} to the magnetization distribution in an equilibrium lattice hopfion, shown in Fig.~1 of Ref.~\onlinecite{M2025}, whose outskirts are visibly more compressed.

\section{Hopfion tails}

Just like in the case of the hopfion lattice~\cite{M2025}, finding the equilibrium hopfion means solving the Euler-Lagrange equation
\begin{equation}
\label{eq:EulerLagrange}
 \frac{\partial F}{\partial e} - \frac{\partial}{\partial t}\frac{\partial F}{\partial e'} = 0,\quad e(0)=0, \quad \lim\limits_{t\rightarrow\infty} e(t) = 1
\end{equation}
for the hopfion profile and simultaneously finding the equilibrium value of $\eta$, minimizing the total energy. Let us first ignore the magnetostatic energy by setting $\mu=0$. Even in this simple case the equation~\eqref{eq:EulerLagrange} is non-linear second order differential equation. It can be easily derived from~\cref{eq:EulerLagrange,eq:Penergy,eq:Pex,eq:Pdm,eq:Pz,eq:Pa} using a computer algebra system, but is too cumbersome to list here.

At small $t\ll1$ we can expand~\eqref{eq:EulerLagrange} into power series in $t$ and recover (in the leading second order) a relation
\begin{align}
\label{eq:smallt}
e''(0) = - 2 (e'(0))^2 .
\end{align}

The hopfion tail is an asymptotic of the profile function $e(t)$ at large $t\rightarrow\infty$, where $e(t) \rightarrow 1$. To find it, let us set $e(t) = 1 - \Delta g(t)$ and expand the eqation~\eqref{eq:EulerLagrange} in powers of $\Delta$. Then, in the leading second order we have
\begin{align*}
 \Delta ^2 \left(g(t) \left(t^2 w^2+2\right)-t \left(t g''(t)+2
   g'(t)\right)\right) = 0,
\end{align*}
where $w=\eta\sqrt{h+q}$. This is a Bessel's-type equation and its solutions can be expressed as a linear combination of the spherical Hankel's functions with imaginary argument $\imath w t$ (proportional to modified spherical Bessel functions) of the first and second kind of the order $-2$, where $\imath=\sqrt{-1}$. The functions of the second kind diverge at $t\rightarrow\infty$ and must be removed from the linear combination. Thus, using Hankel's function asymptotic expansion, we have
\begin{align*}
 \left. e(t)\right|_{t\gg1} \approx 1 - c \frac{e^{-w t} (1 + 1/(w t) + \ldots)}{w t} \approx 1 - c\frac{e^{-w t}}{w t},
\end{align*}
where $c$ is a constant. In a particular case of $h=0$ this coincides with the asymptotic, derived in~\cite{DI79}. Note that, while the static equilibrium hopfion radius $x_0$ in~\cite{DI79} diverges, owing to the Hobart-Derrick theorem~\cite{Hobart1963,*Derrick1964}, the functional form of the asymptotic  solution remains valid.
The hopfion tails decay exponentially and can be considered ``short''. However, there is a catch. The rate of decay is $\propto\sqrt{h+q}$ and can become very small when $h$ comes close to $-q$, which is the stability boundary of the uniformly magnetized state when it is opposite to the applied field. Such situation is not impossible, but to say more it is necessary to study the isolated hopfion's stability and see how close it can approach this phase boundary.

\section{Isolated hopfion stability}

Having established the asymptotics for the hopfion profile function, let us now approximate it with
\begin{align}
\label{eq:trialE}
e(t)= 1-\frac{e^{-t} \left(2 + \left(2 t + 1-\sqrt{3}\right) t\right)}{2 \left(t^3+1\right)},
\end{align}
which satisfies the boundary conditions in~\eqref{eq:EulerLagrange}, has correct derivatives~\eqref{eq:smallt} at $t=0$ and has a correct $1 - e^{-t}/t$ asymptotic behaviour at $t\gg1$. This is not an exact solution to the Euler-Lagrange equation~\eqref{eq:EulerLagrange} anywhere, but just an analytical ansatz, matching both small and large argument asymptotics. Such matching is a very general recipe for building precise analytical approximations. No adjustable parameters in the profile function means that the hopfion energy becomes just a polynomial in $\eta$:
\begin{equation}
  \label{eq:energyTrial}
  \uEE  = \eta \upex + \eta^2 \updm + \eta^3 \left(h \upz + \frac{q}{2} \upa + \mu^2 \upms\right),
\end{equation}
where $\upex\approx67.7287$, $\updm\approx-25.984$, $\upz\approx9.163$, $\upa\approx11.0663$, $\upms\approx2.56789$. These numbers result from substituting~\eqref{eq:trialE} into~\cref{eq:Pex,eq:Pdm,eq:Pz,eq:Pa,eq:Pms} and integrating $\up=\int_0^\infty\uP\ud t$. Minimization of~\eqref{eq:energyTrial} leads to a quadratic equation for $\eta$, but only one of its two solutions corresponds to the energy minimum:
\begin{align*}
\uetaeq = & \frac{d-2 \updm}{6 \uh \upz+3 \uq \upa+6 \mu ^2 \upms},\\
\uEEeq = & -\frac{1}{3}\uetaeq^2 (d+\updm),\\
d = & \sqrt{4 \updm^2-6 \upex \left(2 \uh \upz+\uq \upa+2 \mu ^2 \upms\right)}.
\end{align*}
For $\eta=\uetaeq$ we have $\uEE=\uEEeq$, $\partial\uEE/\partial\eta = 0$ and the second derivative $\partial^2\uEE/\partial\eta^2 = d > 0$. Because the energy~\eqref{eq:Penergy} of the uniform background is zero by construction, equilibrium hopfions correspond to the $\uEEeq<0$ region on the magnetic phase diagram, which is shaded in
\begin{figure}[t]
\begin{center}
\includegraphics[width=\columnwidth]{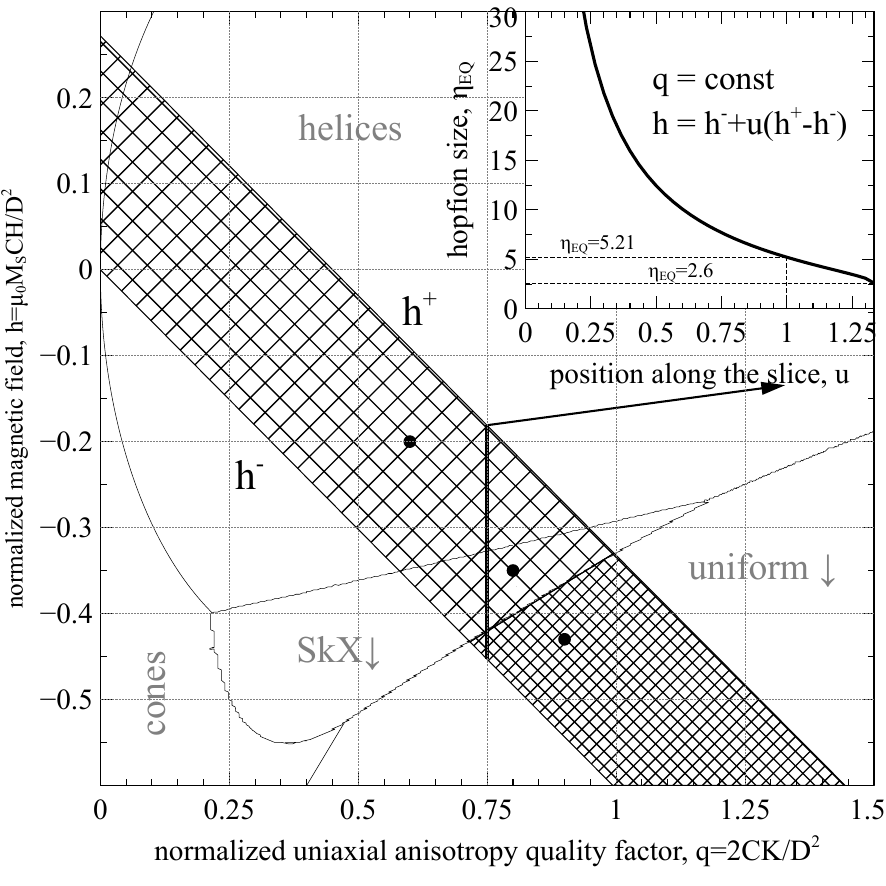}
\end{center}
\caption{\label{fig:phaseD}Phase diagram of a classical helimagnet (with conical, helical, uniform and skyrmion phases) shown in the background; in the shaded region the isolated hopfions have lower energy than their uniform background; in lightly shaded region the uniform background is itself unstable with respect to transition to other non-uniform states of helimagnet; dots show the parameters for which the hopfion profiles were computed numerically and shown in Fig.~\ref{fig:hopfion} lower right; the inset shows the dependence of the equilibrium hopfion radius along a slice of its stability region.}
\end{figure}
Fig.~\ref{fig:phaseD}. This region is contained between the two lines
\begin{align*}
 h^{-} = & \frac{-q \upa-2 \mu ^2 \upms}{2 \upz}, \\ 
 h^{+} = & h^{-} + \frac{\updm^2}{4\upex \upz}.&
\end{align*}
Increase in the magnetostatic parameter $\mu$ causes both equilibrium lines $h^{\pm}$ shift downwards by $-\mu^2\upms/\upz$. Similar overall shift of stability region with increase of $\mu$ was also observed for hopfions in the lattice~\cite{M2024ms}. Note however, that because the trial function~\eqref{eq:trialE} does not account for the influence of the magnetostatic interaction, the present model only works well for small $\mu$, when magnetostatics affects the hopfion size, but not the shape of the profile function. It is also important to remember that there is a stability boundary $h>-q$ of the uniform background itself (when the leftmost energy minimum at the energy plot in  Fig.~\ref{fig:hopfion} top right disappears). It only touches the stability region when $\mu=0$, but cuts into it, as $\mu$ grows. Thus, like in~\cite{M2024ms} for hopfion lattices, the magnetostatic interaction reduces the size of the stability region of isolated hopfions as well.

The hopfion size $\uetaeq$ depends only on the relative position within the hopfion equilibrium strip, but not on the field or anisotropy quality factor directly. Expressing the field within the stability strip via dimensionless parameter $u$ as $h=h^{-} + u (h^{+}-h^{-})$ we can express the hopfion's equilibrium size as
\begin{align*}
 \uetaeq = -\frac{2 \left(\sqrt{4-3 u}+2\right) \upex}{3 u \updm},
\end{align*}
which diverges as the field approaches the lower equilibrium boundary. This dependence is shown on the inset in Fig.~\ref{fig:phaseD}. Note that hopfions remain metastable (although with higher energy than the background) for some small region of $u>1$, until (at larger fields) they eventually get crushed. This limiting hopfion radius is similar to the so-called collapse radius of the magnetic bubbles. Because the $|\uetaeq|>0$ the equilibrium hopfion radius $R_\mathrm{EQ}=\uetaeq R_0= \uetaeq C/D$ diverges when $D\rightarrow0$ in full accordance with the 
Hobart-Derrick theorem~\cite{Hobart1963,*Derrick1964}.

To verify the model, we have also solved the Euler-Lagrange equation~\eqref{eq:EulerLagrange} numerically for some of the points on the phase diagram, marked by black dots in Fig.~\ref{fig:phaseD}. The resulting profiles are shown by dots in Fig.~\ref{fig:hopfion} (lower right). They are almost indistinguishable from the solid lines, showing the rescaled by $\uetaeq$ trial function~\eqref{eq:trialE}.

It is interesting that there seems to be no lower bound for the isolated hopfion stability strip. For hopfions in a lattice, there is a limiting lower stability field $h_{\downarrow}$, which was estimated analytically in~\cite{M2025}. We have performed a similar stability analysis for the present model by considering the trial function with an expanded by $\delta$ small region around $t=t_0$:
\begin{align}
e_\delta(t) = & e(t) + (e(t_0)-e(t)) \theta
   (t-\text{t0}) + \nonumber \\ 
   & (e(t-\delta )-e(t_0)) \theta (t-t_0-\delta ),
   \label{eq:edelta}
\end{align}
where $\theta$ is a unit step function. Setting $t_0=5/8$ corresponds to the expansion by $\delta$ of the domain with $\umz=-1$ between the vortex and antivortex rings of the hopfion, oriented parallel to the (negative) applied field. Yet, the first derivative $\partial\uEE/\partial\delta$, computed from~\eqref{eq:energyTrial} using $\partial\upex/\partial\delta\approx12.4085$,  $\partial\updm/\partial\delta\approx-5.16571$,  $\partial\upz/\partial\delta\approx8.05001$,  $\partial\upa/\partial\delta\approx10.3976$ by substituting~\eqref{eq:edelta} into ~\cref{eq:Pex,eq:Pdm,eq:Pz,eq:Pa}, remains positive in the whole hopfion stability region. It means that the Zeeman's energy gain, achieved by increase of $\delta$, gets overshadowed by the increase in other energy terms and hopfion remains stable. Overall, it seems that the effects of the field and the anisotropy within the hopfion stability strip counter-balance each other even as individually each of these terms grows large. 

Stability of isolated hopfions at negative fields (near the ``knee'' of the saturation-to-saturation hysteresis loop, shown in the inset of the upper right sub-plot in Fig.~\ref{fig:hopfion}) suggests a possibility to obtain stable hopfions on uniform background. This requires material with strong uniaxial anisotropy, so that the background itself remains stable at negative fields. Most of the chiral materials, used in search for magnetic hopfions, like \ce{MnSi} or \ce{FeGe} have very small anisotropy. Also, the magnetostatic interaction destabilizes the hopfions~\cite{M2024ms}. But, as the present analysis shows, they might very well be stable in a material with strong anisotropy and small (compensated) saturation magnetization at negative applied magnetic fields, where their stability region goes outside of the stability regions of other non-uniform phases of the helimagnet, shown as a background in Fig.~\ref{fig:phaseD}.

In positive fields and also in a wider region, marked by light shading in Fig.~\ref{fig:phaseD}, the uniform state itself is unstable with respect to transition into other non-uniform states of helimagnet. The stable hopfions on uniform background can still exist in this region if the background itself is artificially stabilized (e.g. by additional enhanced uniaxial anisotropy layers~\cite{TS2018}). Otherwise the hopfions can only exist on top of other non-uniform background phases. Such complex objects, termed heliknotons, and their dynamics, have indeed been recently observed~\cite{Li2026}. Yet, hopfions on the uniform background are interesting too, because such a background is not a source of pinning, allowing hopfions to move equally well in any direction.

\section{Summary}
We have considered stability of isolated magnetic hopfions on a uniform magnetization background in a classical helimagnet; solved the Euler-Lagrange equation for the hopfion profile asymptotically to determine the shape of its tail; used this knowledge to create a simple and precise analytical model of isolated hopfions; computed their stability range on the phase diagram, which can help directing the search for hopfion-supporting materials.

This work was supported by the Russian Science Foundation under the project RSF
25-22-00076.
\appendix
\section{Energy functions}
\label{app:Pfuncs}
The energy functions can be obtained by substituting~\eqref{eq:ansatz} into~\eqref{eq:energy}, introducing the dimensionless variables and performing the angular integration. This gives:
\begin{align}
 \label{eq:Pex}
 \uPex = & \frac{64 \pi  \left(e^{\prime2} t^2+4 g^2\right)}{3 (1-2 g)^2}\\
 \label{eq:Pdm}
 \uPdm = & \frac{32 \pi  t (4 e g+e^{\prime} (2 g-1) t-2 g)}{3 (1-2 g)^2} \\
 \label{eq:Pz}
 \uPz = & \frac{128 \pi  g^2 t^2}{3 (1-2 g)^2}\\
 \label{eq:Pa}
 \uPa = & \frac{256 \pi  g^2 (4 (g-5) g+5) t^2}{15 (1-2 g)^4},
\end{align}
where $g=e(1-e)$, $e=e(t)$, $e^\prime=e^\prime(t)$.

To evaluate the magnetostatic energy we first compute the density of the volume magnetic charges
\begin{align*}
 \sigma = & \vec{\nabla}\cdot\uvm = \frac{32 g^2 \cos \theta}{(1-2 g)^2 t},
\end{align*}
which, as can be easily seen, neatly factors into the radial and angular part. To make use of this factorization we employ the spherical harmonics expansion
\begin{align*}
%\label{eq:sphericalHarmonics}
 \frac{1}{\left|\uvr -\uvr^\prime\right|}=4\pi\sum\limits_{l=0}^{\infty}\frac{1}{2l+1}\frac{\ur_\mathrm{min}^l}{\ur_\mathrm{max}^{l+1}}\sum\limits_{m=-\infty}^{\infty}Y_l^m(\ut,\uf)\overline{Y_l^m(\ut^\prime,\uf^\prime)}
\end{align*}
for the kernel in the normalized magnetostatic energy integral (includes the factor $1/2$ due to being self-energy)
\begin{align*}
\frac{1}{8\pi}\uiiint \ud^3\uvr\uiiint\ud^3\uvrp\frac{\sigma(\uvr)\sigma(\uvrp)}{|\uvr-\uvrp|},
\end{align*}
where $\ur_\mathrm{min}=\min(\ur,\ur^\prime)$, $\ur_\mathrm{max}=\max(\ur,\ur^\prime)$, $Y_l^m(\ut,\uf)$ are the spherical harmonics and the overline denotes complex conjugate. Taking the angular integrals, we get
\begin{align}
\label{eq:Pms}
 \uPms=& \frac{512 \pi  (e-1)^2 e^2 }{9 (1-2 g)^2 t } \int_{0}^{t} \frac{\tau ^2 g(\tau )^2}{(1-2 g(\tau ))^2} \ud\tau,
\end{align}
where $g(\tau)=e(\tau)[1-e(\tau)]$.

%\bibliography{klm_base}
\end{document}